\documentclass[11pt]{article}
\usepackage{moriond,epsfig}

\def\Journal#1#2#3#4{{#1}{\bf #2}, (#3) #4}

\def\npb{{\em Nucl.~Phys.}~B }

\def\prl{\em Phys.~Rev.~Lett. }
\def\jhep{\em J. High Energy Phys. }
\def\epjc{{\em Eur.~Phys.~J. }{\bf C }} 
\def\prd{{\em Phys.~Rev. }{\bf D }} 
\def\prep{\em Phys.~Rept. }

\begin{document}
\vspace*{4cm}
\title{AZIMUTHAL CORRELATION IN DIHADRON PRODUCTION~\footnote{Talk 
presented at the XXXVIIIth Rencontres de Moriond `QCD and Hadronic Interactions', 
Les Arcs, France, 22-29 March 2002.}}

\author{Alberto GUFFANTI}

\address{Universit\`a di Parma and
INFN, Sezione di Milano, Gruppo Collegato di Parma, Italy\\
\&\\
LPTHE Jussieu, Paris, France\\[10pt]
\epsfig{file=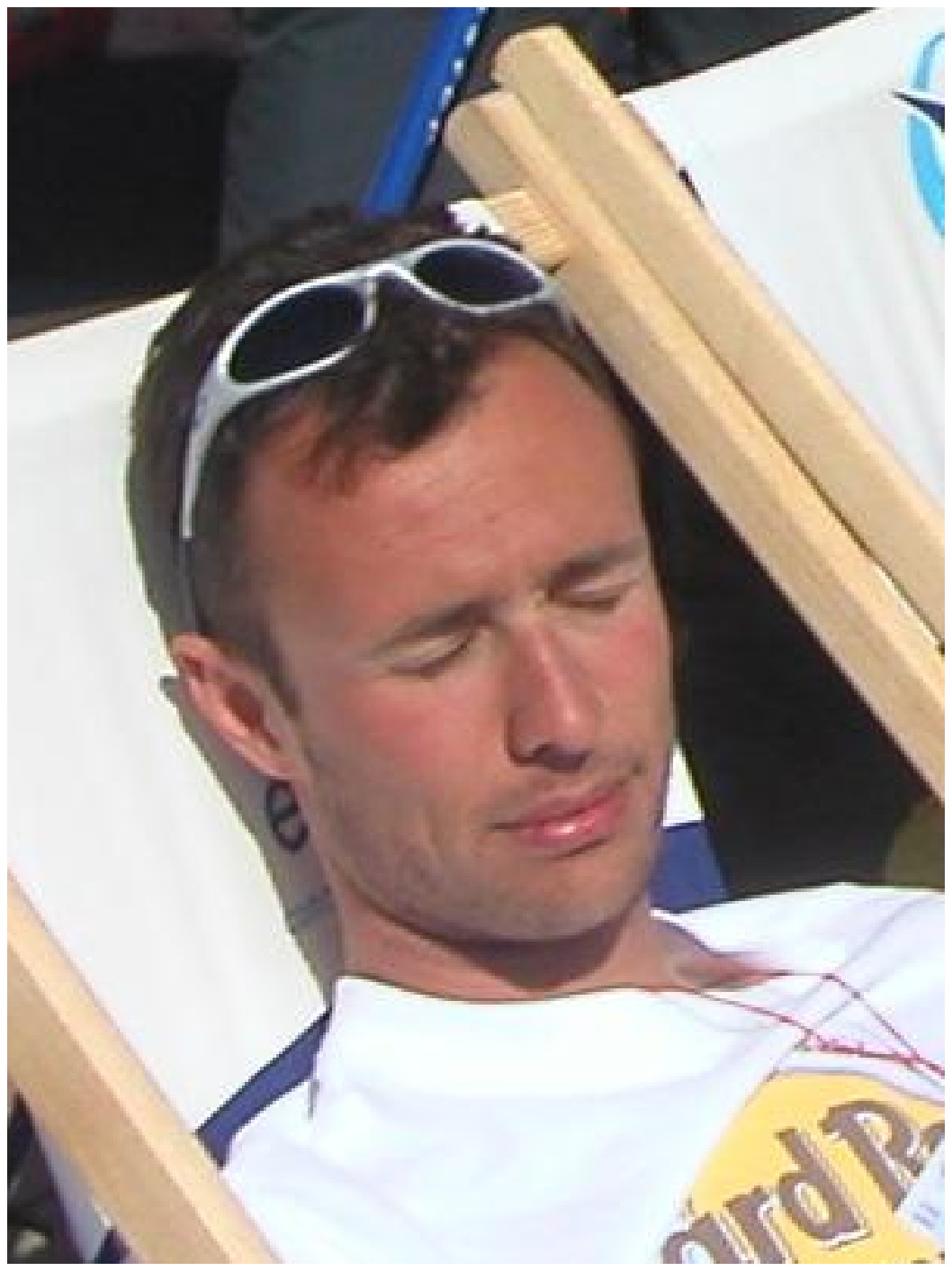,width=35mm}}

\maketitle
\abstracts{
We study the azimuthal correlation of hadron pairs produced in hadron-hadron collisions in the 
framework of perturbative QCD. We concentrate our attention on the `almost back-to-back' region 
where resummation of large logarithms due to soft and/or collinear radiation are shown to play 
an important role. Our aim is to perform perturbative resummation at single logarithmic 
accuracy.}

\section{Introduction}

To our knowledge the azimuthal correlation of a pair of hadrons identified among those 
produced in hadron-hadron collisions has been first proposed as a sensible observable to be
studied in the framework of perturbative QCD (pQCD) at the end of the 70's~\cite{DDT}.
Notably the observable studied was
\begin{equation}
  \label{observable}
  \Sigma\equiv
  \frac{d\sigma}{d p_{c\perp} dy_c d p_{d\perp} dy_d d\chi}\,,
\end{equation}
where $p_{(c,d)\perp}$ and $y_{(c,d)}$ are, respectively, the transverse momenta and the 
rapidities of the outgoing hadrons and $\chi$ is the azimuthal angle, as defined in 
figure 1.\\
Moreover a complete QCD study has been recently carried out for the 
corresponding observable in DIS processes.\cite{azcorrDIS}

From an experimental point of view, the azimuthal correlation in pion-pair production in 
hadronic collision was one of the observables measured by the fixed target E706 \cite{E706} 
experiment at the Fermilab Tevatron.
\begin{figure}[th]
\label{definition-fig}
\begin{minipage}{0.45\textwidth}
\begin{center}
\epsfig{file=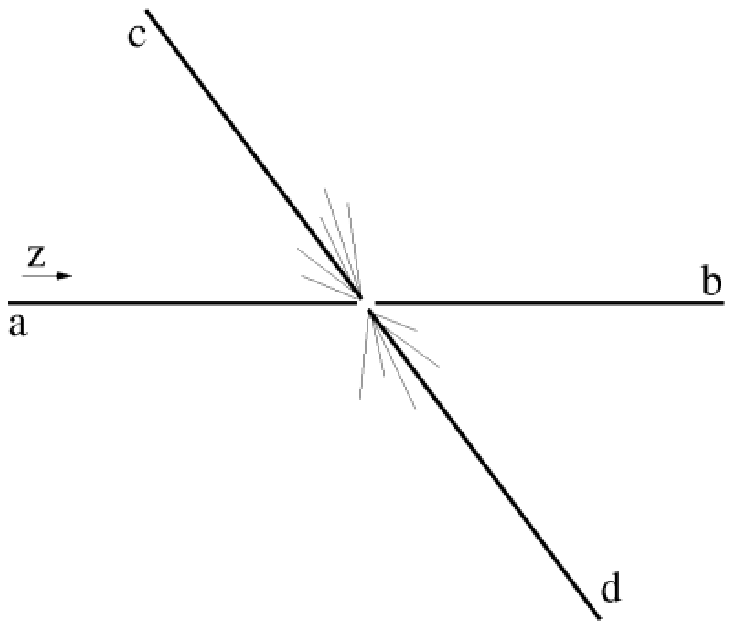, width=5cm}
\end{center}
\end{minipage}
\begin{minipage}{0.45\textwidth}
\begin{center}
\epsfig{file=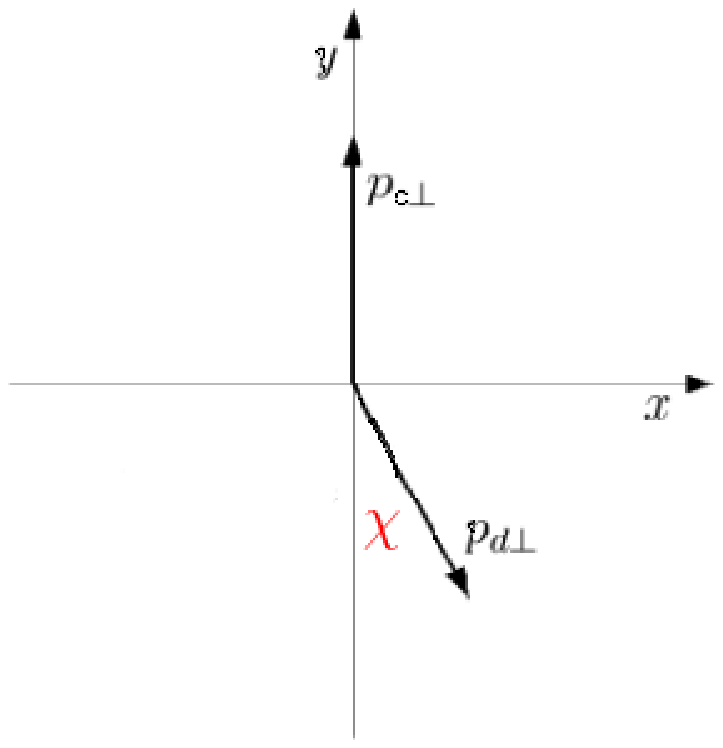, width=5cm}
\end{center}
\end{minipage}
\caption{Definition of the azimuthal angle $\chi$}
\end{figure}

\section{NLO QCD vs. E706 data}

Recently the results of the E706 experiment have been compared to next-to-leading-order (NLO) 
theoretical predictions~\cite{diphox,owens}. As is clearly shown by figure 2, the fixed order 
prediction fails to properly describe the data in the region $\chi\to 0$ (which corresponds to
$\phi\to\pi$ in figure 2).

The reason of this failure is to be ascribed to the fact that for the observable under study 
what we call NLO prediction, i.e. one gluon emission, is, in fact, the first non trivial level 
of approximation. The $2\to 2$ scattering being a planar process, the LO prediction for such 
an observable is simply proportional to $\delta(\chi)$. 

\begin{figure}[hb]
\label{e706vs.data}
 \begin{minipage}{0.45\textwidth}
  \begin{center}
  \epsfig{file=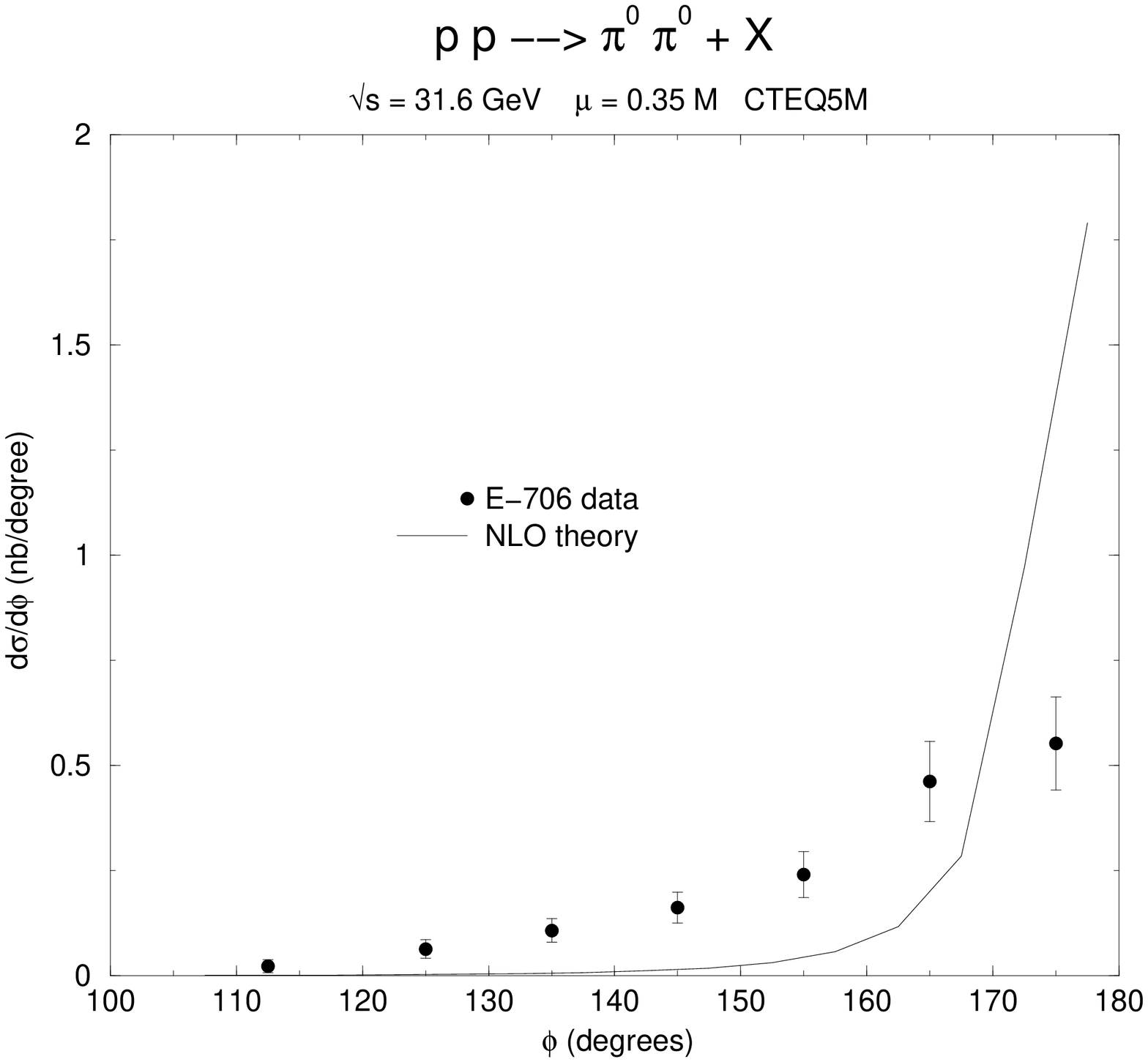, angle=270, width=7cm}
  \end{center}
 \end{minipage}
 \begin{minipage}{0.45\textwidth}
  \begin{center}
  \epsfig{file=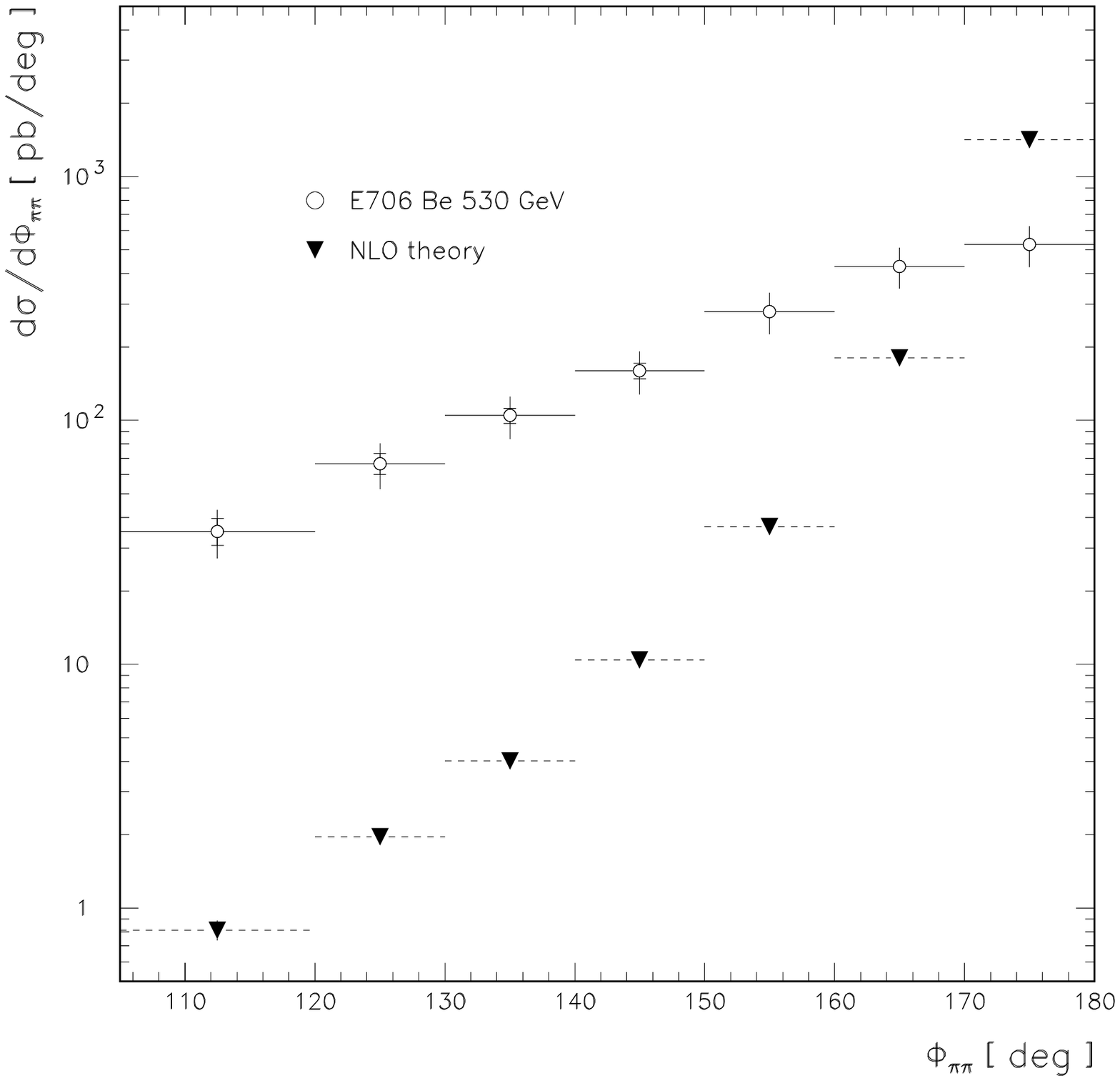, width=7cm}
  \end{center}
 \end{minipage}
  \caption{Comparison of E706 data to the NLO QCD prediction}
\end{figure}

The failure of the NLO prediction in describing the data in the $\chi\to 0$ region can be
seen as an evidence that this region of phase space is dominated by soft and collinear 
multiple emission. In order to obtain a sensible answer we are thus forced to resum logarithmic 
contributions of infrared origin to all orders in perturbation theory.

\section{Resummation}

Up to now  the resummation procedure has been carried out only to double logarithmic (DL) 
accuracy~\cite{DDT}, i.e. taking care of logarithmic contributions of the form 
$\alpha_s^n\ln^{n+1}\chi$ in $\ln\Sigma$, due to gluons which are both soft and collinear. 
This approximation is known to be too crude and largely insufficient to describe the data. 
Anyway this calculation showed the large impact of considering multigluon emission for our 
observable when studying the azimuthal correlation in the $\chi\to 0$ limit.

Our aim is to reach, in the case of this 4-jets~\footnote{We use the term {\em jets} in a
loose acceptation, really meaning hard partons} observable, the same accuracy
attained in 2- and 3-jets ones. This includes resummation of logarithmically enhanced terms 
to single logarithmic (SL) level, which means having under control all terms of the form
$\alpha_s^n\ln^{n+1}\chi$ and $\alpha_s^n\ln^n\chi$ in $\ln\Sigma$, performing the 
matching with the exact NLO prediction and evaluate the leading power corrections.

Resummation to SL accuracy implies taking care of hard collinear emissions and soft large
angle ones. The former contribute to set the scale of parton distribution functions (PDF) 
and fragmentation functions (FF) and the hard scales entering the process. The latter is 
related to the color structure of the underlying event. 

Techniques to deal with soft large angle radiation from 4 hard emitters have been developed 
in the last decade~\cite{Sterman} and imply the {\em refactorization} of the partonic cross 
section into a {\em hard} ($H$) and a {\em soft} ($S$) function, and the use of the corresponding 
renormalization group equation to resum logarithmic contributions to SL level.

The resummation procedure is performed in the impact parameter space ($b$), which is
the Fourier transform of the out-of-scattering-plane momentum, in order to properly take care
of momentum conservation~\cite{PP}
\begin{eqnarray}
  \widetilde{\Sigma}&=&
  \sum_f
  \int\frac{p_\perp db}{\pi}e^{ibp_\perp\chi}
  \prod_{a=1}^2\int_0^1d X_a P_a(X_a, b^{-1})
  \prod_{b=3}^4\int_0^1d Z_b D_b(Z_b, b^{-1})\nonumber\\
  &&\delta\left(\frac{y_c+y_d}{2}-Y\right)
  \delta\left(\frac{p_{c\perp}}{p_\perp}-Z_3\right)
  \delta\left(\frac{p_{d\perp}}{p_\perp}-Z_4\right)
  \frac{1}{16\pi\hat{s}^2}
  \mathrm{Tr}\left\{H_f^{(1)}e^{-{\cal R}_f^\dagger}S_f^{(0)}e^{-{\cal R}_f}\right\}
\end{eqnarray}
where $P_a(X_a, b^{-1})$ and $D_b(Z_b, b^{-1})$ are, respectively, the PDFs and FFs of 
incoming and outgoing partons into corresponding hadrons evaluated at the scale $b^{-1}$.
The sum is over all the partonic subprocesses, $f$,
$H^{(1)}_f$, $S^{(0)}_f$, the hard and soft function, are matrices in color space and depend
on the specific partonic subprocess $f$ and the {\em radiator} ${\cal R}_f$ also a matrix in 
color space, resums all logarithmic contributions up to Single Logarithmic (SL) accuracy.

\section{Conclusions \& Outlook}

In this talk I reported on an ongoing work~\cite{ours} about the study of azimuthal 
correlation of hadron pairs produced in hadronic collision. Such an observable has 
great interest both from an experimental and a theoretical point of view.

Preliminary analytical and numerical results indicate that we should expect, for our observable 
in the region of small $\chi$, a behavior similar to what has been obtained for a similar
observable analyzed in DIS environment~\cite{azcorrDIS} (see Figure 3, for a graph 
showing the behavior of azimuthal correlation in DIS).

If this indeed was the case our prediction would, at least qualitatively, describe the 
behavior of the data in the region of `almost back-to-back' production where present fixed order 
calculation have been shown to be inapplicable.

\begin{figure}[!ht]
\begin{center}
  \epsfig{file=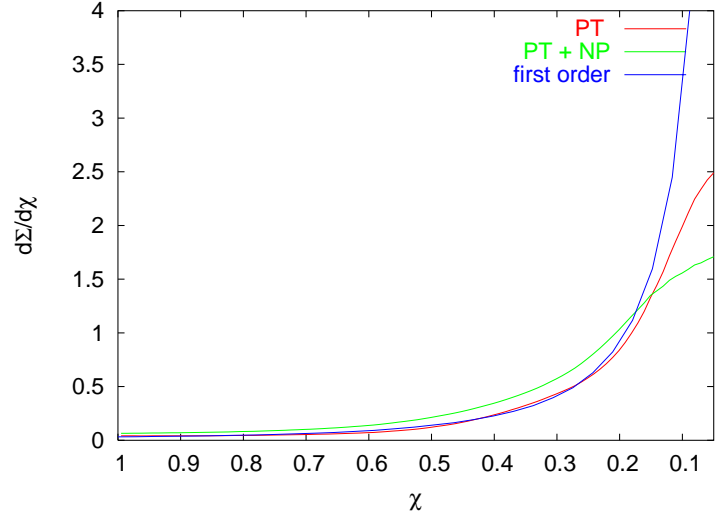, width=10cm}
  \caption{Azimuthal correlation in DIS}
\end{center}
\end{figure}

\section*{Acknowledgments}

I would like to thank the organizers of the Moriond conference for offering me the opportunity
of giving this talk. It is also a pleasure to thank Andrea Banfi for fruitful collaboration on
the subject and Yuri Dokshitzer and Gavin Salam for helpful discussions.

\end{document}